\documentclass[preprint]{emulateapj}

\usepackage{lscape, graphicx} 
\usepackage[final]{pdfpages}
\usepackage{longtable}

\shortauthors{Shi et al.}

\begin{document}

\title{Infrared Spectra and Photometry Of Complete Samples of PG and 2MASS Quasars}

\author{Shi, Yong\altaffilmark{1,2}, Rieke, G. H.\altaffilmark{3}, Ogle, P. M.\altaffilmark{4}, Su, K. Y. L.\altaffilmark{3}, \&  Balog, Z.\altaffilmark{5}}

\altaffiltext{1}{School of Astronomy and Space Science, Nanjing University, Nanjing 210093, China}
\altaffiltext{2}{Key Laboratory of Modern Astronomy and Astrophysics (Nanjing University), Ministry of Education, Nanjing 210093, China}
\altaffiltext{3}{Department Of Astronomy And Steward Observatory, University of Arizona, 933 N Cherry Ave, Tucson, AZ 85721, USA}
\altaffiltext{4}{Infrared Processing and Analysis Center, California Institute of Technology, 1200 East California Boulevard, Pasadena, CA 91125, USA}
\altaffiltext{5}{Max-Planck-Institut f\"ur Astronomie, K\"onigstuhl 17, D-69117 Heidelberg Germany}

\begin{abstract}

As a step toward a  comprehensive overview of the infrared diagnostics
of the central  engines and host galaxies of  quasars at low redshift,
we present  {\it Spitzer} Space Telescope  spectroscopic (5-40 $\mu$m)
and  photometric   (24,  70  and  160  $\mu$m)   measurements  of  all
Palomar-Green (PG) quasars at $z$ $<$ 0.5 and 2MASS quasars at $z$ $<$
0.3. We supplement  these data with {\it IRAS} and  {\it ISO} data and
with {\it Herschel} measurements at 160 $\mu$m. The sample is composed
of  87 optically  selected PG  quasars and  52 near-IR  selected 2MASS
quasars.   Here we present  the data,  measure the  prominent spectral
features,  and  separate emission  due  to  star  formation from  that
emitted by the  dusty circumnuclear torus.  We find  that the mid-IR (
5-30 $\mu$m )  spectral shape for the torus  is largely independent of
quasar IR luminosity with scatter  in the SED shape of $\lesssim$ 0.2 dex.
Except  for the  silicate features,  no large  difference  is observed
between  PG (unobscured  - silicate  emission) and  2MASS  (obscured -
silicate absorption) quasars. Only mild silicate features are observed
in both cases.  When in  emission, the peak wavelength of the silicate
feature  tends  to be  longer  than  9.7  $\mu$m, possibly  indicating
effects on  grain properties near the  AGN.  The IR color  is shown to
correlate  with  the  equivalent   width  of  the  aromatic  features,
indicating that the slope of the quasar mid- to far-IR SED is to first
order driven by  the fraction of radiation from  star formation in the
IR bands.

\end{abstract}                                                    

\section{INTRODUCTION} 

Active  galactic nuclei,  signposts for  accretion  onto super-massive
black  holes (SMBHs), are  now understood  to be  a critical  phase of
massive  galaxy  formation.   In  the local  universe  all  individual
massive galaxies  are found to host  SMBHs at their  centers, with the
SMBH    masses    tightly    correlated   with    galaxy    properties
\citep{Kormendy95, Magorrian98, Gebhardt00, Ferrarese00}.  Probing the
properties of  material located  at different radii  surrounding SMBHs
should  offer important clues  to how accretion  by the
SMBHs influences,  or is  influenced by, the  evolution of  their host
galaxies to maintain this correlation.  It may also solve puzzles such
as how  the angular momentum  of material at  kpc radii is  removed to
feed central engines.  %The infrared data offer %unique diagnostics to
%this  material,  helping  us  to  understand the  dusty  tori,  narrow
%emission %line regions and host galaxies of AGN.

It is believed  that there is a geometrically  thick structure, termed
the dusty torus,  outside the accretion disk in  AGNs. The dusty torus
may physically bridge from the  inner accretion disk to the outer host
galaxy, and  likely plays  a crucial role  in funnelling  the material
into SMBHs.  Significant progress has  been made in probing the torus,
starting from the earliest suggestions  of its existence in the 1980's
to  direct imaging of  it in  a few  nearby AGN  \citep{Jaffe04}.  The
dusty torus is thought to  explain the large apparent diversity of AGN
phenomena. For example, in  the AGN unified model \citep{Antonucci93},
orientation-dependent obscuration in  the torus determines whether the
broad emission lines emerge in the UV/optical spectra and thus whether
an AGN is classified as type 1 or type 2.

The  dust in  the torus  is heated  by UV/optical  radiation  from the
accretion  disk  to  high  temperatures  ($\sim$100-1000  K)  to  emit
dominantly in  the near- to  mid-IR spectral ranges. The  emission by
the torus dust is dominated by a continuum that is featureless, except
for the broad silicate features at  around 9.7 and 18 $\mu$m either in
emission or absorption. Its output is distinct from  the radiation of
star-forming  regions: for  them,  the aromatic  features dominate  at
wavelengths shorter than 15  $\mu$m, while at longer wavelengths there
is  a  strong  continuum  due  to  warm  and  cold  ($<$  60  K)  dust
\citep{Smith07, Stierwalt13}.  In luminous AGN, i.e.  quasars  such as
those  studied here,  does the  torus nearly always  dominate  the integrated
mid-IR emission so  that the observations can be  compared directly to
predictions of torus models.

The  silicate  features are  important  diagnostics  to constrain  the
structure  of the dusty  torus \citep{Siebenmorgen05,  Hao05, Sturm05,
Sturm06, Shi06}. They can be directly compared to predictions of dusty
torus  models, probing the  internal torus  structures \citep{Fritz06,
Honig06,   Nenkova08,  Schartmann08,   Stalevski12,   Feltre12}.   For
example,  \citet{Feltre12} demonstrate how  to discriminate  two broad
types of dusty tori, i.e., smooth vs.  clumpy, through the behavior of
the silicate features and of the overall mid-IR SEDs.  The IR emission
lines also have the advantage  of probing narrow emission line regions
without significant  extinction \citep[e.g.][]{Diamond-Stanic09}. 

The  infrared  data are  also  powerful  in  probing the  host  galaxy
properties, e.g.  the star  formation rate (SFR).  While commonly used
SFR tracers,  such as UV radiation, hydrogen  recombination lines, and
forbidden lines  are contaminated  severely by the  nuclear radiation,
the mid-IR aromatic features  and far-IR photometry are two relatively
uncontaminated    tracers    of     star    formation    in    quasars
\citep[e.g.][]{Shi07,  Netzer07, Lutz08,  Hernan-Caballero09, Hiner09,
Rosario13}.

Studies of 
infrared features are enabled by  the {\it Spitzer}
Space Telescope with a photometric wavelength coverage from 3.6 to 160
$\mu$m and low resolution spectroscopy from 5 to 40 $\mu$m.
To  exploit the  potential  of the  infrared  data, we  carried out  a
program with  {\it Spitzer}  in the last  cryogenic cycle  to complete
spectroscopic (5-40  $\mu$m) and photometric  (24, 70 and  160 $\mu$m)
observations of two  samples of luminous AGNs at  low redshift, namely
the   unobscured  optically   selected   Palomar-Green  (PG)   quasars
\citep{Schmidt83,  Boroson92}  and  obscured  near-IR  selected  2MASS
quasars  \citep{Cutri01, Smith02}.   By focusing  on quasars,  we will
probe the phase  where SMBHs grow most rapidly  and where their output
stands out most clearly from that of their host galaxies.

In this paper we present  the observations and extract measurements of
aromatic and  silicate features  and continuum luminosities  for these
two samples.   We use  this information to  determine the SFRs  of the
host galaxies and to show that  the values from the aromatic bands and
far  infrared  emission  are   consistent.   We  present  the  sample,
observations, and  data reductions in  \S~\ref{sample_selection}.  The
spectral  decomposition  is  detailed  in  \S~\ref{spec_decomp}.   The
results  are  presented  in  \S~\ref{results}.   Our  conclusions  are
presented \S~\ref{conclusions}. Throughout this paper, we assume
$H_{0}$=70    km    s$^{-1}$    Mpc$^{-1}$,    $\Omega_{0}$=0.3    and
$\Omega_{\Lambda}$=0.7.

\section{Samples, Observations And Data Reduction}\label{sample_selection}

\subsection{Samples}

For  this  study  we  included   all  87  objects  of  the  PG  sample
\citep{Schmidt83, Boroson92} at $z$ $<$  0.5 and all 52 objects of the
2MASS sample  \citep{Cutri01, Smith02}  at $z$ $<$  0.3, as  listed in
Table~\ref{sample}.   The  PG  quasars  are defined by an  average
limiting $B$  band magnitude of 16.16, blue $U$-$B$ color
($<$ -0.44), and dominant star-like appearance.  All these objects show
broad emission lines, and thus  are classified as type 1 quasars.  Due
to  the  large photographic  magnitude  errors  and  the simple  color
selection, the  PG sample is  incomplete \citep[e.g.][]{Goldschmidt92,
Jester05},  but  the  incompleteness  is independent  of  the  optical
magnitude and color \citep{Jester05}, indicating that the PG sample is
still representative  of bright optically  selected quasars.  Compared
to PG  quasars, the 2MASS  quasars represent a redder  population with
$J$-$K_{s}$ $>$ 2 (compared with a typical value of $J$-$K_{s}$ $\sim$
1.5  for  PG  quasars),  but have  similar  $K_{s}$-band  luminosities
($M_{K_{s}}$ $<$  -23) \citep{Smith02}.   Unlike PG quasars  the 2MASS
sample includes  objects with narrow, intermediate  and broad emission
lines. The 2MASS  sample is increasingly incomplete at  $K_{s}$ $>$ 13
\citep{Cutri01}.

\subsection{Observations And Data Reduction}\label{observation}

We carried out 25.1 hr  of {\it Spitzer} observations (PID: 50196; PI:
G.  Rieke  ) to  complete the IRS  spectroscopic and  MIPS photometric
data for the  PG/2MASS sample.  The program included  all objects that
did not  have archived data from  previous cycles. As  a result, there
are low resolution infrared  spectra (5-40 $\mu$m) and MIPS photometry
at  24, 70  and 160  $\mu$m for  the entire  sample.\footnote { The  previous IRS
spectroscopic programs were by  J. Houck (PID: 4), G. Rieke (PID:
36), F. Low  (PID: 40), M. Werner (PID: 61),  S. Gallagher (PID: 148),
R.  Siebenmorgen (PID:  193), D.  Lutz (PID:  323), S.  Veilleux (PID:
485),  ,  A.  Stockton  (PID:  14067),  A.  Wehrle  (PID:  14991)  and
 particularly  by P. Ogle (PID:  11451). The previous MIPS programs were
led by  F. Low  (PID: 49),  G. Rieke (PID:  82, 30306,  40053, 50507),
M.  Werner (PID:  86), M.  Bondi (PID:  3327), M.  Bondi  (PID: 3327),
Z.  Shang (PID:  20084), A.  Marscher  (PID: 20496),  A. Wehrle  (PID:
30785), G. Fazio (PID: 30860).} All the broad-band photometry is listed in
 Table~\ref{tab_broadband}, while the IRS
data are available in Table~\ref{data}.

The  spectra were  obtained using  the standard  IRS staring  mode. To
reduce them, we drew from  the archive the basic calibrated data (BCD)
as products  of {\it Spitzer}  Science Center data  reduction pipeline
version  S18.7,  which provided  cosmic  ray  removal, replacement  of
saturated   pixels,   droop    correction,   subtraction   of   darks,
linearization correction,  and stray light and  flat field correction.
The   post-pipeline    processing   of   the   BCD    was   based   on
IRSCLEAN\footnote{http://irsa.ipac.caltech.edu/data/SPITZER/docs/dataanalysistools/tools/irsclean/}
and
SPICE\footnote{http://irsa.ipac.caltech.edu/data/SPITZER/docs/dataanalysistools/tools/spice/}.
IRSCLEAN was used to create bad pixel masks for the BCD image. The sky
background for  each module within  a given order was  then subtracted
using  the  image obtained with the  same module  in  a  different order.   For
observations with only  one order, the image at  one slit position was
used as sky background for  the image at another slit position.  SPICE
was employed  to extract the spectra  from these background-subtracted
images.  We used the optimal  extraction mode to increase the ratio of
signal to noise, since our targets are point sources.  The mismatch in
the signal between short-low (5-14 $\mu$m) and long-low (14-40 $\mu$m)
modules was removed by scaling  the short-low spectrum to the long-low
one.   The whole spectrum  was further  scaled to  the MIPS  24 $\mu$m
photometry by  comparing to synthetic photometry of  the spectra. 
The  flux  difference between  the  short-low  and  long-low, and  the
difference  between the MIPS  24 $\mu$m  photometry and  IRS synthetic
flux at  this wavelength,  could be  caused by the  errors in  the flux
calibration of  the MIPS and IRS  instruments, and the pointing  errors with
short-low and  long-low, not necessarily by  extended emission outside
the  IRS slit.   As listed  in Table~\ref{tab_broadband},  the scaling
factor  from short-low  to  long-low has  a  median value  of 1.0  and a 
standard deviation of  0.4, and the factor from  long-low synthetic 24
$\mu$m  to MIPS  24  $\mu$m flux  has a  median  value of  1.08 and  a
standard deviation  of 0.18. The  almost unity conversion  factor with 
scatter comparable to the flux calibration uncertainties indicates that 
missed  extended  emission   is  not   important  for   the  IRS
observations. In addition,  visual checks of MIPS 24  $\mu$m images do
not find any source with significant extended emission.

MIPS observations were made with  the small field photometry mode. The
data reduction and photometry  measurements were carried out with Data
Analysis  Tool v3.1  and  redMIPS v1.1  \citep{Gordon05}  by the  MIPS
Instrument Team. For the 160 $\mu$m photometry, we have visually
checked individual  images and removed  those where the  photometry is
affected  by  close companions  or  structure  in  the sky  (e.g.,  IR
cirrus).  In  total  98\%, 91\%  and  32\%  of  the sample  have  MIPS
detections above 3$\sigma$ at 24, 70 and 160 $\mu$m, respectively.

To increase  the fraction  of objects with  160 $\mu$m  detections, we
searched for measurements in  the {\it Herschel} Space  Telescope data
archive. All  the PG quasars have  been observed with {\it  Herschel} (PI:
L. Ho)  but no archived data  were found for the 2MASS  quasars.  These data
were reduced to images as described in \citet{Balog13}. First, bad  and
saturated pixels were flagged.  The response  was calibrated and corrected for
flat fields.  A high-pass filter was applied to eliminate the 1/f
noise  of  the  detector.   Aperture  photometry  was
carried out on the reduced images as in  \citet{Balog13}. For objects with both  {\it Spitzer} 160
$\mu$m and {\it Herschel} 160 $\mu$m detections, the latter is used because of
(generally) higher signal-to-noise and greater freedom from structured sky emission.

%%Additional infrared photometry  observed with previous satellites is also
%%included.  The  Infrared Astronomical Satellite  (IRAS) photometry was
%%retrieved  using a 50  arcsec  search  radius in  the  IRAS faint  source
%%catalog.   The Infrared  Space Observatory  (ISO) data  were collected
%%through    NED\footnote{http://ned.ipac.caltech.edu/}.     All    data
%%including spectra and photometry are listed in Table~\ref{data}.

\section{Spectral Decompositions}\label{spec_decomp}

\subsection{Fitting spectral features}

\begin{figure*}
\epsscale{0.8}\plotone{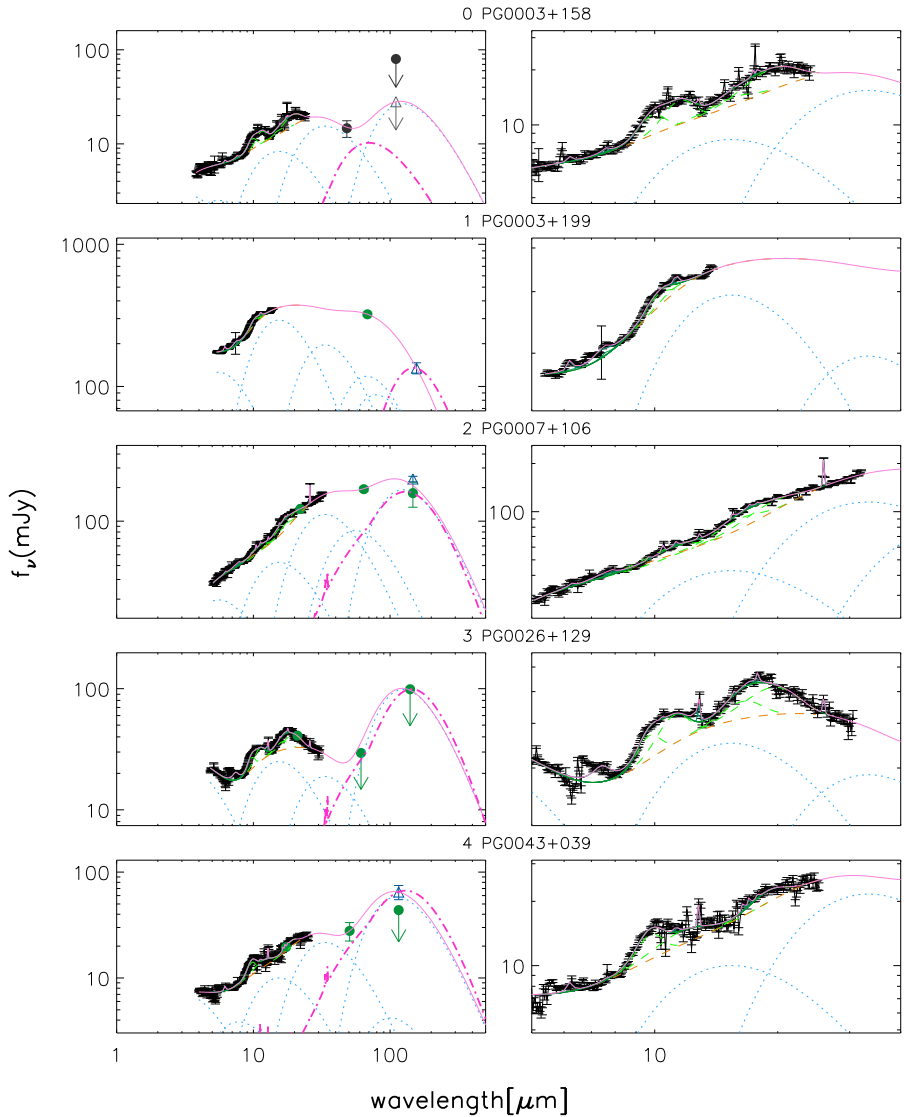}
\caption{\label{IR_SED} The rest-frame  IRS spectra (black points) and
broad-band  photometry (green filled  circles).  For  each object,  the left
panel shows the full SEDs while the right panel zooms into the 5 to 40
$\mu$m spectral range. The pink solid lines are the fitted curves as a
sum of  all individual fitted profiles  including black-body, silicate
features, aromatic features and emission lines.  The blue dotted lines
indicate the fitted individual  black-body emission  the sum  of  which is
shown as  yellow dashed  line.  The green  dashed lines  represent the
fitted silicate  feature profiles. The pink dotted-dashed line is the best-fit
star-forming template derived from the library of \citet{Rieke09}.}
\end{figure*}

The  infrared spectra of  quasars are  rich in  features. As  shown in
Fig.~\ref{IR_SED}, they usually contain  strong hot dust continuum emission, cold
dust emission, silicate emission or absorption  features at 9.7
and 18  $\mu$m, aromatic  features mainly at  6.2, 7.7, 8.6,  11.3 and
12.1 $\mu$m,  plus atomic fine-structure and molecular hydrogen emission lines.  

\begin{figure}
\plotone{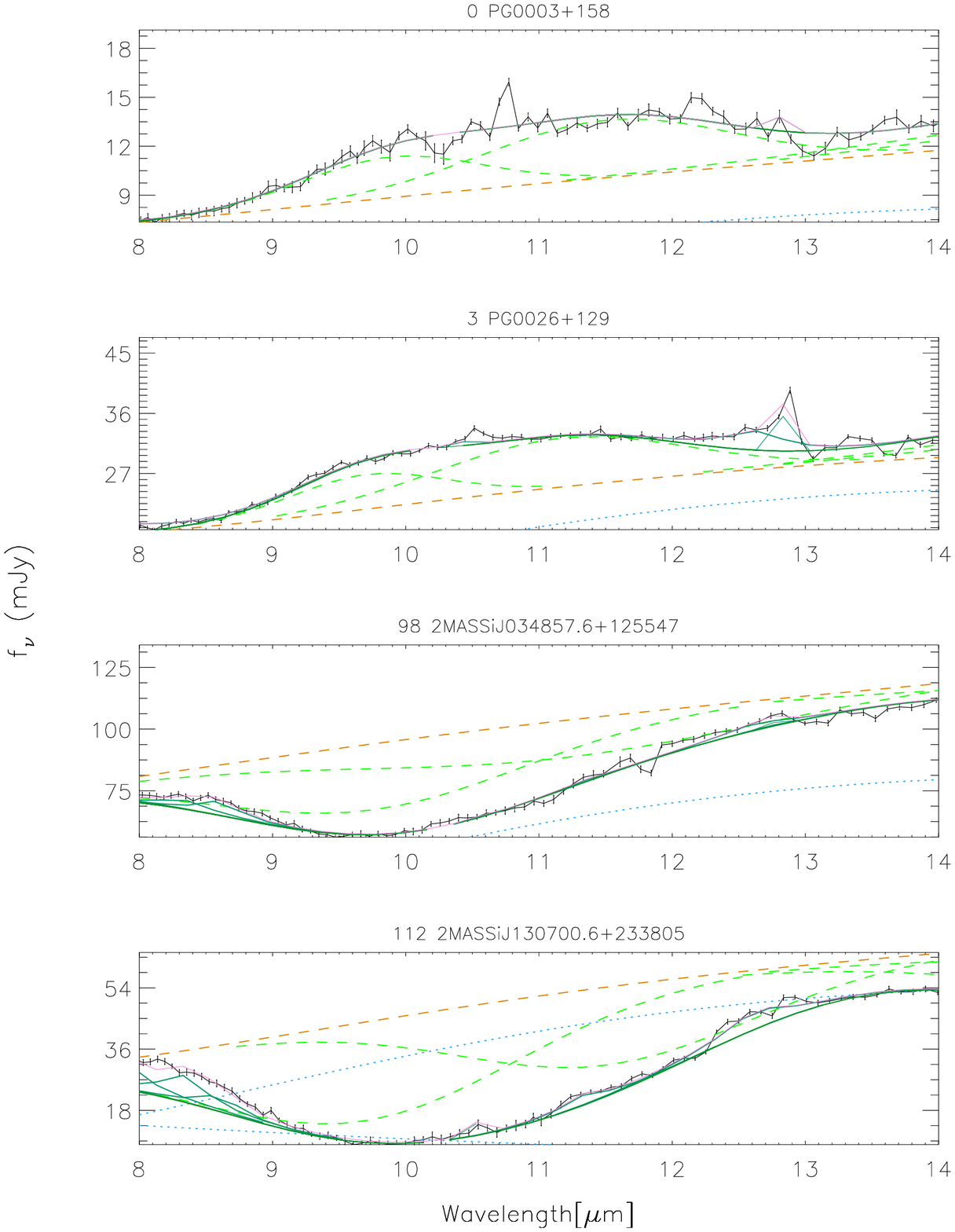}
\caption{\label{IRSED_silicate} Examples of two Gaussian fitting for the 9.7 $\mu$m silicate feature. 
All line styles are the same as the Figure 1.}
\end{figure}

Although these features can be measured individually, we prefer to fit
them simultaneously  so that blends  can be separated.  Our  method is
based  on  a  simple  physical  model,  similar  to  the  IDL  program
PAHFIT.pro  that was  developed  by \citet{Smith07}  to decompose  the
spectra of normal galaxies.  The  dust continuum is modelled through a
combination of a series of black-bodies at fixed temperatures of 1000,
580, 415,  210, 97,  58, 48,  41, 32, 29  and 26  K while  letting the
normalization be free.  The  highest temperature is chosen  to fit
the  hottest dust  emission peaking  around 3  $\mu$m, and  the lowest
temperature is  limited by our SED  coverage that in  general does not
extend  beyond 200  $\mu$m.  This  set  of temperatures  was based  on
several test runs, and includes  criteria such as the peak wavelengths
of the black-bodies  should not lie within spectral  regions of strong
and broad silicate and aromatic features.  The emission at wavelengths
longer   than  200  $\mu$m   is  usually   described  by   a  modified
black-body. However, as  we have little $>$ 200  $\mu$m photometry and
are not  fitting features  in this spectral  range, we  simply adopted
black-body spectra. Higher-fidelity fits to these long wavelengths are
discussed in  the following section.  Each silicate  feature either in
emission or absorption is modelled through two Gaussian functions: 1.)
for  the  9.7 $\mu$m  silicate  feature,  central  wavelengths of  two
Gaussians  were set  to  be 10.0  and  11.5 $\mu$m  that  can vary  by
$\pm$3\%.  The standard deviations (i.e., widths) were set  to be 8\% and 10\% of the
central  wavelengths, respectively.   The fractional Gaussian  widths were
allowed to vary from -2\% to +4\%. The normalizations of the two
Gaussians are free, with the initial value set to be half of
the  difference between  the  observed flux  at  the Gaussian  central
wavelength  and  the one  at  8$\mu$m;  and  2.)  for  the 18  $\mu$m
feature,  two   Gaussians  were  placed  at  16.0   and  19.0  $\mu$m,
respectively, with all other parameters treated  similarly to those for the 9.7 $\mu$m
feature; the initial value of the  normalization is set to
be half  of the difference between  the observed flux  at the Gaussian
central    wavelength    and    the    one   at    15$\mu$m.    
Figure~\ref{IRSED_silicate} shows examples of the two-Gaussian fitting for
the 9.7  silicate feature.   A single Gaussian  function was  used to
model the emission lines including H$_2$ S(3) 9.67 $\mu$m, [SIV] 10.52
$\mu$m,  [NeII] 12.81  $\mu$m, [NeIII]15.56  $\mu$m, H$_2$  S(1) 17.03
$\mu$m and [OIV] 25.91 $\mu$m. The  line width of the Gaussian was set by
the spectral  resolution at a given wavelength but  allowed to  vary by
$\pm$20\%,  while the  normalization was  totally  free.  As
detailed in  \citet{Smith07}, the aromatic features  were described by
Drude profiles with fixed centers and widths but free normalizations.

The  fitting  results  from  minimizing  $\chi^{2}$  were  in  general
excellent  under visual  inspection. However,  for about  20\%  of the
sample, the  broad silicate emission  features appeared to  mimic dust
black-body emission. In these cases, the initial value of the silicate
feature strength needs  to be relatively accurate and  was obtained by
spline interpolation or visual  estimate.  We used a general criterion
that  the  dust continuum  underlying  the  silicate emission  feature
between 9 and 20 $\mu$m should change gradually and monotonically.

The results of the  SED decompositions are shown in Fig.~\ref{IR_SED}.
With the  fitted profiles, various features are  quantified and listed
in Table~\ref{data_feature}.  The silicate feature strength is defined
as ln($f^{\rm  peak}$/$f^{\rm peak}_{\rm cont}$)  where $f^{\rm peak}$
is  the flux  density at  the wavelength  where the  silicate emission
feature peaks or the silicate absorption feature shows a minimum based
on the two  fitted Gaussian profiles for the  individual features, and
$f^{\rm peak}_{\rm  cont}$ is the  continuum flux density at  the same
wavelength.   The  continuum luminosities  at  given wavelengths  were
measured as  the average of the  spectra over a 2  $\mu$m range around
the   central   wavelengths.     All   the   listed   errors   in
Table~\ref{data_feature}  only  consider  the  photon  noise  in  the
spectrum. For the silicate feature intensity, due to the difficulty in
differentiating the broad feature  from the underlying continuum, there
are  additional   systematic  uncertainties   that  could  be   up  to
0.1-0.2.  The peak  wavelength  can also  suffer  systematic errors  of
$\sim$ 1  $\mu$m for the  9.7 $\mu$m feature  and 2 $\mu$m for  the 18
$\mu$m silicate feature.

\subsection{Decomposing  SEDs   into  star-forming  and   dusty  torus
components}

The  above  spectral  decomposition  aims to  measure  intensities  of
different spectral features. In this  section we decompose the full IR
SED into emission from star formation and the radiation from the dusty
torus.  For  an initial  reconnaissance, we assume  the emission  at a
far-IR  wavelength   or  an  aromatic  feature   comes  entirely  from
star-forming  regions  and   then  select  the  star-forming  template
\citep{Rieke09}  that  yields the  closest  observed  fluxes. We  have
derived  such  estimates in  three  ways:  1.)   11.3 $\mu$m  aromatic
features;  2.) MIPS 70  $\mu$m photometry;  and 3.)  MIPS or  PACS 160
$\mu$m photometry.

Although  the aromatic  features  in quasar  spectra  arise from  star
formation   regions  in   the  host   galaxies  \citep[e.g.][]{Shi07},
observations  of  the  central  regions  ($\sim$0.5  kpc)  of  Seyfert
galaxies indicate that the nuclear radiation may suppress the aromatic
features at 6.2,  7.7 and 8.6 $\mu$m, but not  the 11.3 $\mu$m feature
\citep{Diamond-Stanic10, Esquej14}.  We  convert the aromatic flux for
this   feature   to   SFR   using  the   star-forming   templates   of
\citet{Rieke09}.   An issue with  this method  is the  large intrinsic
scatter between  aromatic fluxes and  star formation rates as  seen in
non-AGN star-forming galaxies \citep{Smith07, Calzetti07}.

Far infrared luminosity  is generally considered to be  a reliable SFR
indicator for galaxies. For each broad-band photometry measurement, we
choose the star-forming template with monochromatic luminosity closest
to the  observation, normalize  it to the  measured flux  density, and
estimate the  FIR luminosity from this normalized  template.  The risk
for this approach  is that the far-IR emission  may be contaminated by
cold dust  in the dusty  torus.  After choosing the  best star-forming
template, including bands  not affected by cold ISM  dust, we measured
the SFRs  based on  the 24 $\mu$m  emission of the  template following
\citet{Rieke09}.

\begin{figure*}
\plotone{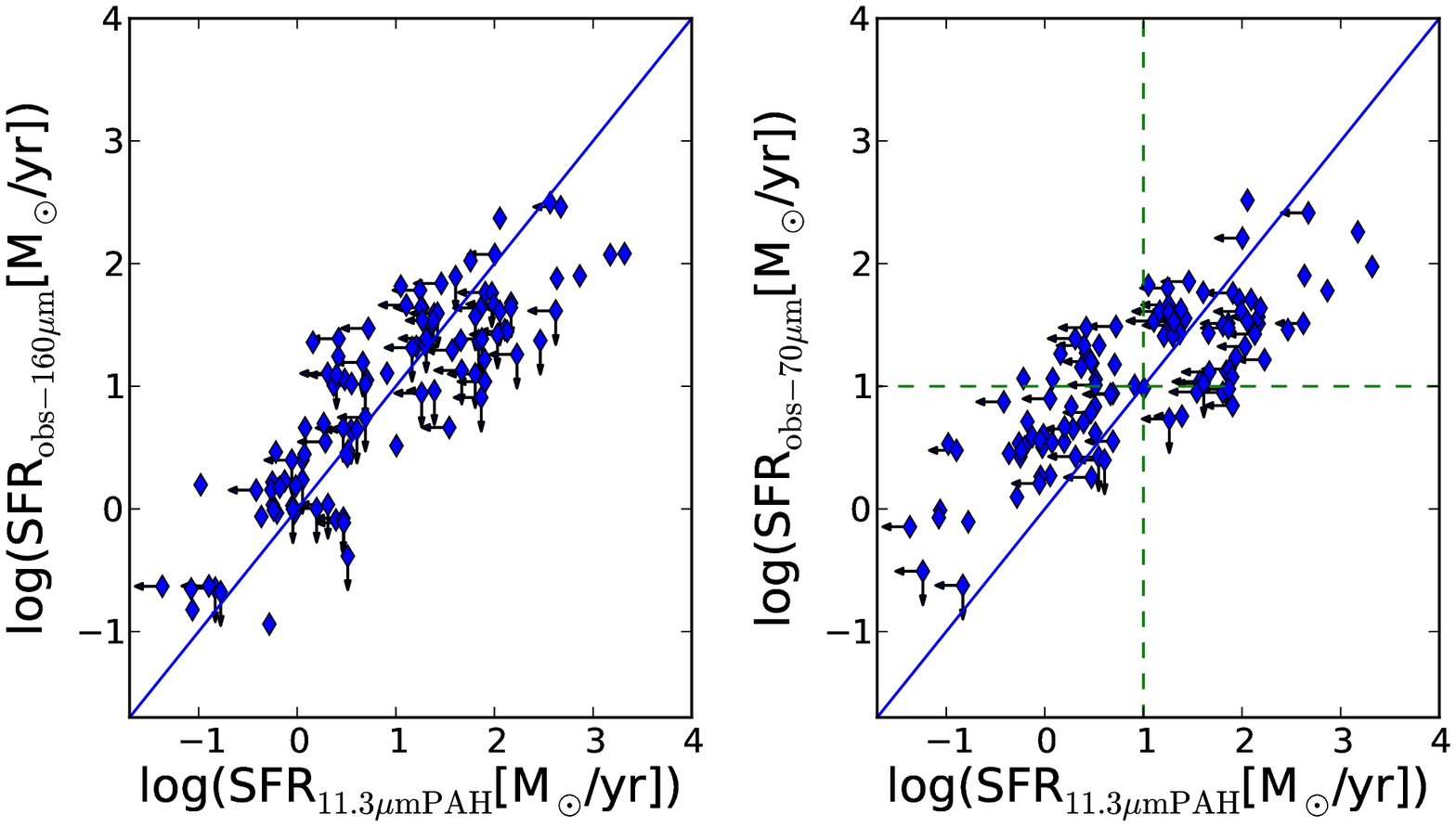}
\caption{\label{LSFR_LSFR}    Comparisons   between    different   SFR
estimates: the left panel shows the comparison between estimates 
from the 160  $\mu$m and aromatic feature measurements;
the right panel shows the comparison between 70 $\mu$m-based
and aromatic-feature-based SFR estimates.  }
\end{figure*}

We evaluate  the contribution from the  dusty torus to the  70 and 160
$\mu$m  emission  by  comparing  SFRs  from these  two  bands  to  the
aromatic- based  SFRs, since  the aromatic emission  should arise
mainly from star forming  regions. Although aromatic features could be
excited by B-type stars \citep{Peeters04, Vega10}, for our sample with
relatively   high   far-IR    luminosity   (mostly   above   10$^{10}$
L$_{\odot}$), such  a contribution should be negligible.  As shown in
Figure~\ref{LSFR_LSFR},   no  systematic   offset   is  seen   between
160$\mu$m-based  SFRs  and  the   SFRs  from  aromatic  features.  The
situation for the 70$\mu$m-based SFRs is more complex. For SFRs larger
than  10  M$_{\odot}$/yr  from  the  aromatic  feature,  there  is  no
systematic offset.   For lower SFRs,  the estimates from  the 70$\mu$m
measurement tend to be high.  This behavior may indicate contamination
of the 70 $\mu$m photometry  (typically at rest frame $\sim$ 50 $\mu$m
for our  sample) by emission from  the dusty torus in  cases where the
SFR is low.

The above  three methods demonstrate  the overall consistency  of SFRs
estimated   from  monochromatic  photometry   or  a   single  aromatic
feature. Therefore, it  is permissible to improve the  accuracy in the
SFR estimates by combining the three methods through template fitting.
For  objects with  160 $\mu$m  detections, a  first-guess star-forming
template is  chosen to have the  closest 160 $\mu$m  luminosity to the
observation.   If this  template, however,  produces higher  70 $\mu$m
luminosity than is observed, we discard it and instead choose the next
colder  template in  the library  of \citet{Rieke09}.   Sometimes this
process  needs to  be  repeated  until the  best  template is  chosen.
Similarly,   for  objects   with  70   $\mu$m  detections,   the  best
star-forming  template is  the  one  that has  the  closest 70  $\mu$m
luminosity  to the  observation.  But,  if the  template  produces 160
$\mu$m  output above the  observed upper-limit,  a hotter  template is
chosen.  As  indicated in Figure~\ref{LSFR_LSFR}, for  70 $\mu$m based
SFRs smaller than 10 M$_{\odot}$/yr, we reduce the SFR estimate by 0.3
dex to account statistically for the torus emission.  For objects with
aromatic features, the best-fit template is required to not exceed the
70 and 160 $\mu$m upper-limits; if it does it is renormalized to lower
its luminosity and  make it consistent. The estimated  SFRs are listed
in   Table~\ref{data_feature}.    Both   quasar   samples   have   SFR
measurements for as many as 94\% of the members, with upper limits for
the rest.   We assigned a systematic  error of 0.3 dex  for 160 $\mu$m
based SFRs  as well as 70  $\mu$m based SFRs  above 10 M$_{\odot}$/yr,
and 0.5 dex  for 70 $\mu$m based SFRs below  10 M$_{\odot}$/yr as well
as   aromatic   based   SFRs,     based   on   the   scatter   of
Figure~\ref{LSFR_LSFR}.    Fig.~\ref{IR_SED}   shows   the   best-fit
star-forming template for each quasar.  The residual after subtracting
the star-forming template  is adopted as the radiation  from the dusty
torus.

\section{Results}\label{results}

\subsection{Composite Quasar SEDs and Scatter About Them}

\begin{figure}
\epsscale{1.0}\plotone{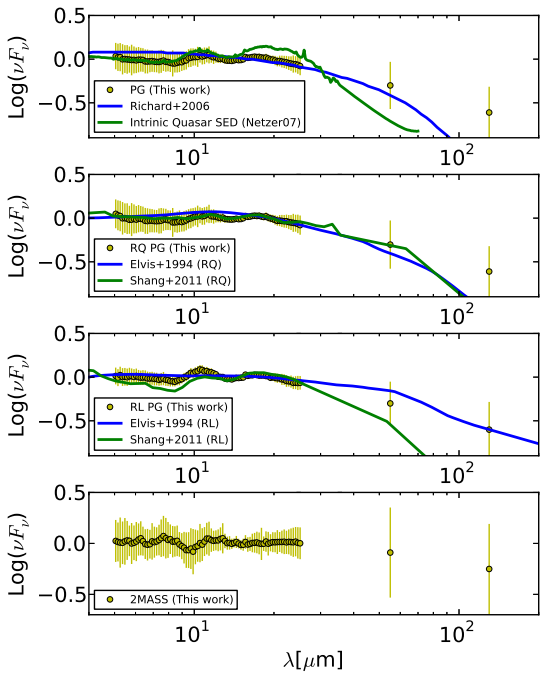}
\caption{\label{COMP_SPEC} From top to bottom, filled circles indicate composite 
median SEDs of all PG quasars, radio quiet PG quasars, 
radio loud PG quasars and all 2MASS quasars, respectively, where the error bar is shown
at 1-$\sigma$ level. The comparison SEDs include those
by \citet{Elvis94}, \citet{Richards06}, \citet{Netzer07} and \citet{Shang11}.  }
\end{figure}

Figure~\ref{COMP_SPEC}  shows   the  median  composite   spectra  with
1-$\sigma$ scatter  for the  PG and 2MASS  samples normalized  at 14-16
$\mu$m.  The composite spectrum from 5 to 25 $\mu$m is derived by
taking  the median  of the  spectroscopic data  projected to  a common
wavelength grid with a spectral  resolution of 50.  We took the median
of all the 70 $\mu$m photometry with the rest-frame wavelength from 40
to 70 $\mu$m  as the composite flux density at 55  $\mu$m, and that of
160 $\mu$m data with the  rest-frame wavelength from 100 to 160 $\mu$m
as the  composite flux density at  130 $\mu$m. We also  constructed composite
spectra for the radio quite  and radio loud sub-samples of PG objects,
where  the  radio  loudness  data are  from  \citet{Kellermann89},  as
indicated in  Table~\ref{COMP_SPEC}.  All  the PG objects  have mid-IR
spectroscopic  measurements while  90\% and  76\% of  the  sample have
detections  at 70  and 160  $\mu$m, respectively.   95\% of  the 2MASS
sample   have  mid-IR   spectroscopic  measurements   and   70  $\mu$m
detections, with  160 $\mu$m detections for 33\%.   The median spectra
presented  in  Fig.~\ref{COMP_SPEC} should  be  representative of  the
whole PG sample up to rest-frame 130 $\mu$m and of the 2MASS sample to
rest-frame 60  $\mu$m, while the  rest-frame 130 $\mu$m data  point of
the  2MASS median  spectrum should  be biased  by the  far-IR luminous
objects  in the  sample.  There  are  some differences  in the  median
spectra  between  the 2MASS  and  PG  samples.   While the  PG  median
spectrum  shows   silicate  emission,  the  2MASS   one  has  silicate
absorption, consistent  with the near-IR selected  2MASS quasars being
viewed on average  more edge-on.  In addition, the  2MASS spectrum has
much larger EWs of the  aromatic features compared to the PG spectrum,
as  well as  elevated rest-frame  60 $\mu$m  emission relative  to the
mid-IR. This behavior indicates more active star formation relative to
the   nuclear  SMBH   luminosity   in  2MASS   quasar  host   galaxies
\citep{Shi07}.

Figure  3  compares the  SEDs  we have  determined  with  a number  of
alternatives.    \citet{Elvis94}  presented   composite  SEDs   of  47
non-blazar quasars, largely based on  members of the PG sample.  Their
IR data are based on ground-based N  and Q bands (10 and 20 $\mu$m) as
well  as IRAS broad-band  photometry (12,  25, 60  \& 100  $\mu$m). As
shown in Figure~\ref{COMP_SPEC}, their radio-quiet composite SED shows
an overall  similar shape to that  of radio-quiet PG  quasars from our
work  below  rest-frame  60  $\mu$m  but  is  significantly  lower  at
rest-frame 130  $\mu$m. However, their data do  not include photometry
at wavelengths  longer than observed-frame  100 $\mu$m and the  SED at
longer wavelengths is based on extrapolation. On the other hand, their
radio-loud quasar composite SED  is similar to ours between rest-frame
5   and  130  $\mu$m.    \citet{Richards06}  compiled   {\it  Spitzer}
broad-band photometry of 259 SDSS-selected optically bright quasars at
3.6, 4.5, 5.8,  8.0, 24 and 70 $\mu$m. Their  sample covers a redshift
range from $z$=0  up to $z$=4.  Compared to  our composite spectrum of
all  PG quasars,  their SED  shows  an overall  similar shape  between
rest-frame 5 and 30 $\mu$m but lacks silicate emission features. Above
rest-frame 100 $\mu$m, their SED drops faster than ours.  Again, their
observations do not  contain far-IR data above observed  70 $\mu$m and
the  shape at  longer  wavelengths is  based  on extrapolation.  
\citet{Netzer07} constructed the intrinsic  quasar SED of 8 PG quasars
after  removing the  star-forming  emission. Their  SED resembles  our
composite spectrum for the whole PG sample up to 30 $\mu$m above which
theirs  drops  much  faster,  which  is expected  as  the  star-forming
radiation starts  to dominate the infrared emission  above 30 $\mu$m.
\citet{Shang11} constructed  a composite  SED of 85  optically bright,
non-blazar   quasars  that   are   selected  heterogeneously.    Their
radio-quiet  composite SED  is  quite similar  to  our radio-quiet  PG
spectrum between rest-frame 5 and  30 $\mu$m but drops faster longward
of rest-frame 55  $\mu$m, which is also the  case when comparing their
radio-loud composite SED to our  radio-loud PG one.  The comparison of
160  $\mu$m photometry  for about  20 objects  in common  reveals that
their measurements are systematically lower than ours by a factor of a
few. Since  our independent measurements at this  wavelength from {\it
Spitzer} and {\it Herschel} agree  much more closely than this factor,
we believe our measurements are reliable.

\begin{figure}
\epsscale{1.0}\plotone{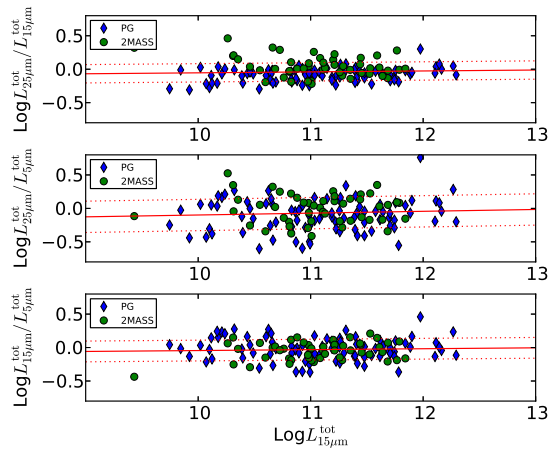}
\caption{\label{lum_lum} The mid-IR color of 25$\mu$m/15$\mu$m, 25$\mu$m/5$\mu$m, 15$\mu$m/5$\mu$m as a function of
the 15 $\mu$m luminosity. All luminosities are monochromatic emission averaged over 2 $\mu$m width of the central
wavelength. The best linear fits are shown as solid lines with 1-$\sigma$ scatter as the dotted line. The fitting results are  also listed in Table~\ref{tab_lum_lum}. }
\end{figure}

\begin{figure*}
\epsscale{1.0}\plotone{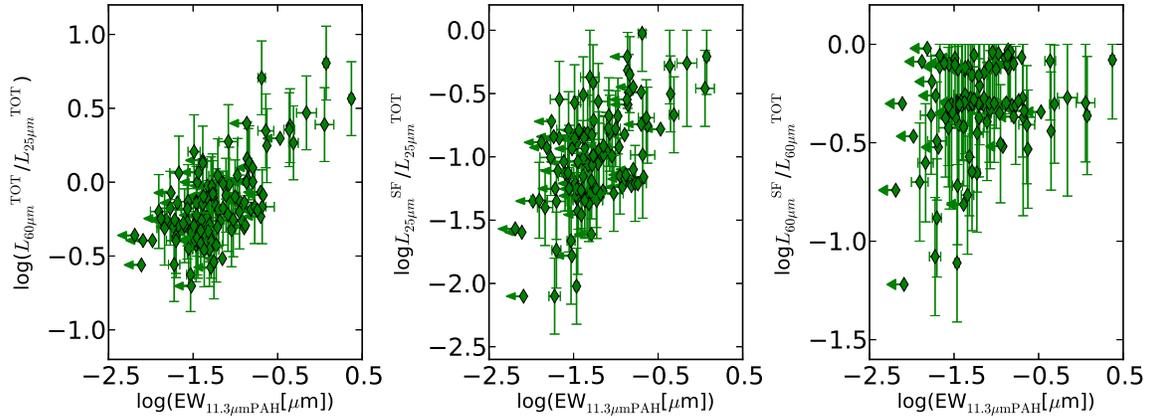}
\caption{\label{SFfrac_EWPAH}   {\it  Left   Panel:}   The  IR   color
($L_{60{\mu}m}^{\rm  TOT}$/$L_{25{\mu}m}^{\rm TOT}$)  as a  function of
the 11.3 $\mu$m aromatic feature equivalent width. {\it Middle Panel:}
The fraction of  25 $\mu$m emission from star  formation as a function
of  the 11.3  $\mu$m  aromatic feature  equivalent  width. {\it  Right
Panel:} The fraction  of 60  $\mu$m emission from  star formation  as a
function of the 11.3 $\mu$m aromatic feature equivalent width. }
\end{figure*}

To illustrate the scatter of the mid-IR spectra among the whole PG and
2MASS samples, we present  the relationships between infrared color as
a    function   of   15    $\mu$m   monochromatic    luminosities   in
Figure~\ref{lum_lum}  and  list   the  best-fit  linear  relations  in
Table~\ref{tab_lum_lum}.   Almost   zero  slopes  for   all  of  these
relationships indicate that the mid-IR  spectral shape of PG and 2MASS
quasars  are roughly  independent of  quasar mid-IR  luminosities.  In
addition, the scatter of the relationship is quite small, ranging from
0.14  to 0.23  dex. This  suggests that  the composite  mid-IR spectra
derived  above  are  good  representatives of  quasar  mid-IR  spectra
independent  of quasar  IR  luminosities and  quasar types.   However,
these   spectra  include  combinations   of  SMBH-   and  SF-generated
components, as  would also be  the case for the  previously determined
quasar   spectral    templates   \citep[e.g.][]{Elvis94,   Richards06,
Shang11}.   Obtaining an  intrinsic  quasar SED  template  in the  far
infrared requires subtracting the SF contribution.

\subsection{Contribution of Star Formation to the SED}\label{hot_dust_sec}

\begin{figure}
\plotone{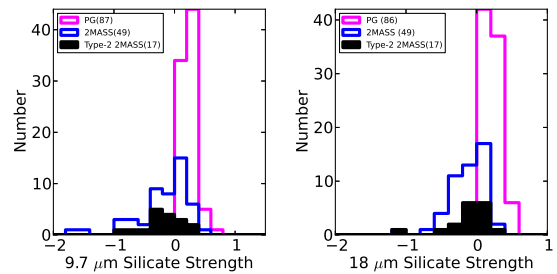}
\caption{\label{silicate_strength}   The  distributions   of  silicate
feature strengths for PG and 2MASS quasars. The  9.7 $\mu$m feature is
shown in  the left  panel and the  18 $\mu$m  feature is in  the right
panel.}
\end{figure}

The  left  panel  of  Figure~\ref{SFfrac_EWPAH}  shows  the  IR  color
log($L_{60{\mu}m}/L_{25{\mu}m}$) as  a function of the EW  of the 11.3
$\mu$m aromatic feature.   This color varies from -0.8  to 0.8 for the
whole quasar sample, but it clusters  in a smaller range from -0.5 to
0.25.  The  EW$_{\rm 11.3{\mu}mPAH}$ for  the sample ranges  from $<$0.003
$\mu$m to  2.3 $\mu$m. As  shown in the  figure, the two  quantities are
roughly     related     to     each     other,     with    increasing
log($L_{60{\mu}m}/L_{25{\mu}m}$)      at      increasing      EW$_{\rm
11.3{\mu}mPAH}$,  which   indicates  that  the   rest-frame  IR  color
log($L_{60{\mu}m}/L_{25{\mu}m}$) is  to the first order  driven by the
relative  brightness of star-forming  regions and  the dusty  tori.  A
relationship between IR color $f_{30{\mu}m}/L_{15{\mu}m}$ and EW$_{\rm
7.7{\mu}mPAH}$ is  also observed  in the sample  of \citet{Veilleux09}
that is composed of PG  quasars and ULIRGs. However, such correlations
are  not  always  observed  in  other  samples  of  galaxies  and  AGN
\citep{Wu10, Stierwalt13}.   \citet{Wu10} investigated the IR
SEDs of about 300 galaxies and  AGN selected to be brighter than 5 mJy
at 24 $\mu$m. They found  their objects are divided into two categories,
one       with       low       EW$_{\rm       6.2{\mu}mPAH}$       and
log($f_{70{\mu}m}/f_{24{\mu}m}$),  and   another  with  high  EW$_{\rm
6.2{\mu}mPAH}$  and   log($f_{70{\mu}m}/f_{24{\mu}m}$)  but  with  no
apparent    correlation    between    EW$_{\rm   6.2{\mu}mPAH}$    and
log($f_{70{\mu}m}/f_{24{\mu}m}$) within the category.   \citet{Stierwalt13} mainly focused
on  the IR  SEDs  of LIRGs  and  did not  see  a relationship  between
$f_{30{\mu}m}/f_{15{\mu}m}$  and EW$_{\rm  6.2{\mu}mPAH}$  either.  As
mentioned by \citet{Stierwalt13}, the IR slope is not only affected by
the star-forming contribution but is also affected by the obscuration,
which could  cause a steep IR  spectrum even at low  EW$_{\rm PAH}$ if
the energy source  is heavily buried. Our quasar  samples lack heavily
buried  sources, as  indicated by  the distributions  of  the silicate
feature  in Figure~\ref{silicate_strength}. The  mid-IR slope  is then
mainly  driven   by the star-forming   contribution  and  thus   shows  a
correlation with the EW$_{\rm PAH}$.

We  further investigated the  star-forming contribution  at individual
wavelengths (25 or 60 $\mu$m) and their trend with the EW$_{\rm PAH}$.
As  shown  in  the  middle  panel  of  Figure~\ref{SFfrac_EWPAH},  the
star-forming fraction  at rest-frame 25 $\mu$m is  usually below 30\%,
indicating the  emission at this wavelength is  generally dominated by
the radiation from  the dusty torus. This fraction  is correlated with
the EW$_{\rm 11.3{\mu}mPAH}$,  as might be expected because  the SF is
not the dominant contributor, and both the quasar and star-forming
SEDs do not change significantly  from source to source. On the other
hand, as  shown in the  right panel of  Figure~\ref{SFfrac_EWPAH}, the
rest-frame 60  $\mu$m radiation is  dominated by star  formation, with
the star-forming  fraction mostly above  30\%. This fraction  does not
show an apparent relation  with the EW$_{\rm 11.3{\mu}mPAH}$, again as
might be expected since the two indicators are mainly affected by
different  processes,  SF  dominating  the  60  $\mu$m  while EW$_{\rm 11.3{\mu}mPAH}$  
is a measure of the fraction of the output at this wavelength due to star formation to that from nuclear
accretion, since the latter dominates the 11.3 $\mu$m continuum.

\subsection{ Silicate Features} \label{silicate_sec}

Figure~\ref{silicate_strength} shows the  distributions of the 9.7 and
18  $\mu$m silicate  feature strengths  including  non-detections. The
non-detected  features  are  weak  with 3-$\sigma$  upper  limits  for
ln($f^{\rm peak}$/$f^{\rm peak}_{\rm cont}$)  between -0.3 and 0.3. As
shown in  the figure, PG  quasars mainly show emission  features while
the majority  of 2MASS quasars  have features in absorption,  which is
consistent  with the  expectations from  the unified  model  where the
dusty tori of  the unobscured quasars are viewed  face-on and those of
obscured quasars  are seen  edge-on.  However,  there is  no good
relation between the silicate  feature and quasar optical type; e.g.,
many type 2 quasars show silicate features in emission.
As indicated in the figure, the feature strength is moderate, with the
majority in the range  of ln($f^{\rm peak}$/$f^{\rm peak}_{\rm cont}$)
= -0.5  to 0.5.   The lack  of strong ($>$0.5  or $<$-0.5)  9.7 $\mu$m
silicate  emission features  is  consistent with  expectations if  the
dusty  torus  is clumpy  instead  of  smooth \citep{Fritz06,  Honig06,
Nenkova08, Schartmann08, Stalevski12, Feltre12}.

\begin{figure}
\plotone{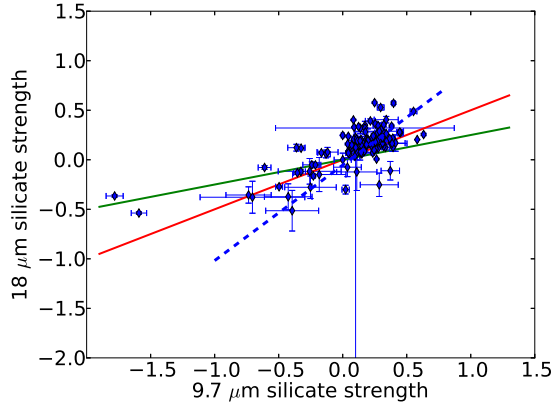}
\caption{\label{silicate97_silicate98} The  18 $\mu$m silicate feature
strength  vs.   9.7 $\mu$m  feature  strength.   Symbols indicate  the
measurements of PG and 2MASS objects where the dashed line is the best linear
fit.  The red and green areas  are the predictions for clumpy and smooth
dusty tori, respectively \citep{Feltre12}.}
\end{figure}

 Figure~\ref{silicate97_silicate98} shows the relationship between
two   silicate   feature   strengths   for   sources   with   EW$_{\rm
11.3{\mu}mPAH}$ $<$ 0.2. The two features are roughly correlated, with
a Pearson value of 0.6.  A  regression fit to the data points gives
$S_{18{\mu}m}$=0.96$S_{9.7{\mu}m}$-0.05 and 1-$\sigma$ scatter of 0.83.
The relatively  large scatter partly arises from  the uncertainties in
determining    the     underlying    continuum.     As     shown    in
Figure~\ref{silicate_strength},  the small  dynamic range  of  the two
feature  strengths   will  also  limit  the  accuracy   in  the  slope
measurement.   In Figure~\ref{silicate97_silicate98}, we  overlay some
predictions of clumpy and  smooth models \citep{Feltre12}.  The clumpy
model  produces a  slope closer  to the  observations than the smooth model.  In a dust geometry composed of clumps, the
sides of  individual clouds facing  the central nuclei are  heated and
produce silicate  emission features while the  absorption features are
imposed on  the output  from the dark  sides.  In an  inclined viewing
angle toward the  AGN, the foreground clouds partially  block the view
of the  background ones.  Therefore,  a majority of clouds  are viewed
from  the  dark  sides  and  silicate  absorption  features  dominate,
although some fraction of clouds  are seen from their bright sides and
the  associated   silicate  emission  reduces  the   strength  of  the
absorption  trough. As  the 9.7  $\mu$m feature  has a  larger optical
depth than the 18 $\mu$m  one, implying a stronger emission feature at
9.7  $\mu$m  from  the  bright  side  of the  cloud,  the  9.7  $\mu$m
absorption  feature  is  reduced  more  so that  a  steeper  slope  is
produced.

\begin{figure}
\plotone{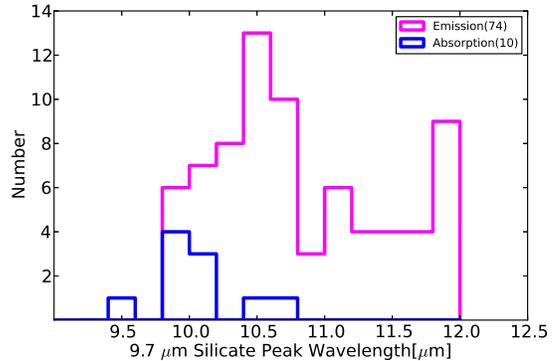}
\caption{\label{silicate_pk_wave}  The distribution of the peak wavelength of the ``9.7 
$\mu$m" silicate feature for all cases with S/N $>$ 5.}
\end{figure}

Figure~\ref{silicate_pk_wave}  shows  the  distribution  of  the  peak
wavelengths of 9.7  $\mu$m features with S/N $>$  5.  Almost all ``9.7
$\mu$m"  features peak  in  emission at  wavelengths  longer than  9.7
$\mu$m, the  peak location of  silicate features observed  in Galactic
sources or  star-forming galaxies  \citep{Smith10}. Some  of them
even peak at the longest wavelength (12 $\mu$m) that is allowed by our
fit,  indicating that they could  peak at even longer  wavelength.  In
contrast, those in  absorption show minima at the  expected 9.7 $\mu$m
wavelength.  The offsets in  the peak wavelengths of emission features
in   quasars  have   been   observed  previously   in  small   samples
\citep{Siebenmorgen05, Hao05,  Sturm05, Sturm06, Shi06}.   None of the
existing dusty torus models that  adopt Milky Way ISM dust composition
predict such offsets \citep{Fritz06, Honig06, Nenkova08, Schartmann08,
Stalevski12, Feltre12}.   We have used  the library of  embedded young
stellar object  SEDs from  Whitney (2014) to  test whether  the offset
could arise from radiative transfer effects, but find that our fitting
procedure generally centers the  silicate features near 9.7 $\mu$m for
them. We  conclude that the offset  is intrinsic to  the quasars. Such
differences between  silicate absorption and  emission features likely
indicate that the physical properties  of the emitting dust grains are
subject  to  modifications  due  to  exposure to  the  strong  nuclear
radiation, while ``normal"  dust grains in the host  galaxies or outer
cold  edges of  the  dusty  tori are  responsible  for the  absorption
features.  It has  been shown that for individual  cases, the silicate
emission feature in  quasars can be well reproduced  by adopting large
size dust grains or  modifying the dust compositions \citep{Molster03,
Smith10}.

\begin{figure}
\epsscale{1.0}\plotone{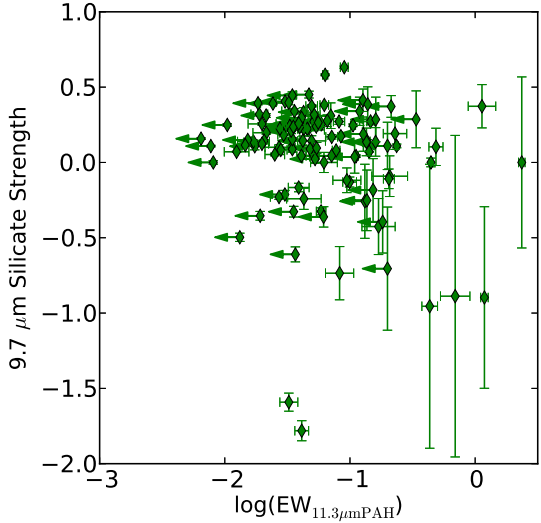}
\caption{\label{SIL_EW} The 9.7 $\mu$m silicate feature strength as a function of
the 11.3 $\mu$m aromatic feature equivalent width. }
\end{figure}

 Figure~\ref{SIL_EW}  shows the 9.7  $\mu$m silicate  feature strength
vs.   the EW  of the  11.3 $\mu$m  aromatic feature.   No relationship
between the two  is observed, which is also seen  in other AGN samples
\citep{Hernan-Caballero11}.  Unlike LIRGs  and  ULIRGs \citep{Spoon07,
Stierwalt13},  our  quasar sample  does  not  show  any deep  silicate
features  (strength $<$-1.5)  at any  EW$_{\rm  11.3{\mu}mPAH}$, while
LIRGs  and ULIRGs contain  a branch  in this  plane that  shows deeper
silicate features at  decreasing EW$_{\rm 11.3{\mu}mPAH}$, pointing to
a population of deeply buried sources.

\section{Conclusions}\label{conclusions}

We  report   mid-infrared  spectroscopy  and   mid-  and  far-infrared
photometry of  the PG and 2MASS  samples of quasars.  We analyze these
data  by   fitting  for  spectral   features  due  to   aromatics  and
silicates.  The  broadband infrared  SEDs  of  both  quasar types  are
similar. However,  the PG quasars  tend to have silicates  in emission
and the  2MASS ones  have them  in absorption. We  show that  the star
formation  rates estimated  from  the aromatic  features  and the  far
infrared luminosities of these quasars are consistent. Emission due to
star formation  in the  host galaxies dominates  the outputs  of these
systems at 160 $\mu$m, but at  70$\mu$m and low SFRs it is likely that
the quasar  contaminates the star-forming SED  significantly. The peak
wavelength of the '10$\mu$m' silicate  feature tends to be longer than
observed in the general interstellar medium, suggesting that the grain
properties are modified by proximity to the AGNs.

\section{Acknowledgment}

We thank the  anonymous referee for helpful comments  that improve the
paper significantly.   Y.S.  acknowledges  support for this  work from
Natural Science Foundation of China under the grant of 11373021 and by
the   Strategic   Priority  Research   Program   ``The  Emergence   of
Cosmological Structures''  of the  Chinese Academy of  Sciences, Grant
No.  XDB09000000.

%%%%%%%Physical Properties Of PG quasars Included In This Sample 

\LongTables
\begin{deluxetable*}{ccccccccccc}
\tabletypesize{\scriptsize}
\tablecolumns{8}
\tablecaption{\label{sample} Physical Properties Of Our Sample}
\tablewidth{0pt}
\tablehead{ 
  \colhead{Number}  & \colhead{Name}  &  \colhead{Position(J2000)} & \colhead{$z$} & \colhead{$m_{B}$} & \colhead{$m_{K}$} & \colhead{Type} \\
  \colhead{(1)}     & \colhead{(2)}   &  \colhead{(3)}      & \colhead{(4)} & \colhead{(5)}     & \colhead{(6)} & \colhead{(7)} 
}
\startdata
1   & PG0003+158                 & 00h05m59.24s  +16d09m49.0s &  0.450    &  15.960         &                &  1.0           \\
2   & PG0003+199                 & 00h06m19.52s  +20d12m10.5s &  0.025    &  13.750         &                &  1.0           \\
3   & PG0007+106                 & 00h10m31.01s  +10d58m29.5s &  0.089    &  16.110         &                &  1.0           \\
4   & PG0026+129                 & 00h29m13.60s  +13d16m03.0s &  0.142    &  14.950         &                &  1.0           \\
5   & PG0043+039                 & 00h45m47.27s  +04d10m24.4s &  0.384    &  15.880         &                &  1.0           \\
6   & PG0049+171                 & 00h51m54.80s  +17d25m58.4s &  0.064    &  15.880         &                &  1.0           \\
7   & PG0050+124                 & 00h53m34.94s  +12d41m36.2s &  0.061    &  14.390         &                &  1.0           \\
8   & PG0052+251                 & 00h54m52.10s  +25d25m38.0s &  0.155    &  15.420         &                &  1.0           \\
\enddata 
\tablecomments{ Col. (2): the name of the target; Col. (3): the position of the target; Col. (4): redshift; Col. (5):
B-band magnitude of PG quasars in Vega system \citep[see][]{Schmidt83}; Col. (6): $K_{s}$ magnitude of 2MASS quasars in
internal 2MASS magnitude system\footnote{http://www.ipac.caltech.edu/2mass/releases/allsky/doc/sec6$\_$4a.html}. \\
 The complete version of this table is in the electronic edition of the Journal. Col. (7): The optical type of emission line where 1.0
for broad emission line, 2.0 for narrow emission line and values between 1.0 and 2.0 for intermediate types. }
\end{deluxetable*}

%%%%%%%%%%%%%%%%%%%%%%%%%%%%%%%%%%%%%%%%%%%%%%%%

\LongTables
\begin{deluxetable}{ccccccccccccccccc}
\tabletypesize{\scriptsize}
\tablecolumns{8}
\tablecaption{\label{tab_broadband} BROAD-BAND PHOTOMETRY OF THE SAMPLE}
\tablewidth{0pt}
\tablehead{ 
\colhead{Number}  & \colhead{Name}  &  \colhead{$f_{\rm Spitzer-24{\mu}m}$} &  \colhead{$f_{\rm Spitzer-70{\mu}m}$} & \colhead{$f_{\rm Spitzer-160{\mu}m}$} & \colhead{$f_{\rm Herschel-160{\mu}m}$} & \colhead{C$_{\rm IRS-SL}^{\rm IRS-LL}$}  & \colhead{C$_{\rm IRS}^{\rm MIPS}$}\\ 
\colhead{}        & \colhead{}      &  \colhead{[mJy]}     & \colhead{[mJy]}     & \colhead{[mJy]} & \colhead{[mJy]} & \colhead{} & \colhead{} \\
\colhead{(1)}     & \colhead{(2)}   &  \colhead{(3)}  & \colhead{(4)}       & \colhead{(5)}   & \colhead{(6)}  & \colhead{(7)}  & \colhead{(8)}
}
\startdata
  1 & PG0003+158                     & 24.69$\pm$0.38  & 21.26$\pm$4.29  & $<$115.88       & $<$40.68        & 1.31            & 0.99           \\
  2 & PG0003+199                     & -9.00$\pm$1.00  & 329.80$\pm$11.0 &                 & 138.06$\pm$12.3 &                 & 1.77           \\
  3 & PG0007+106                     & 139.90$\pm$0.29 & 210.80$\pm$9.31 & 194.40$\pm$49.1 & 260.19$\pm$14.8 & 1.00            & 0.92           \\
  4 & PG0026+129                     & 46.82$\pm$0.37  & $<$33.67        & $<$112.96       &                 & 1.29            & 1.13           \\
  5 & PG0043+039                     & 27.08$\pm$0.41  & 38.53$\pm$7.63  & $<$60.59        & 89.71$\pm$13.76 & 0.95            & 1.18           \\
  6 & PG0049+171                     & 17.46$\pm$0.49  & $<$14.48        & $<$163.63       & $<$40.56        & 1.00            & 1.06           \\
  7 & PG0050+124                     & 953.40$\pm$0.48 & 2084.00$\pm$11. & 1612.00$\pm$34. & 1884.34$\pm$17. & 0.96            & 1.06           \\
  8 & PG0052+251                     & 71.19$\pm$0.33  & 82.07$\pm$8.68  & $<$150.31       & 155.53$\pm$16.2 & 0.98            & 0.99           \\
\enddata
\tablecomments{ The  factor in the column (7) is used to 
multiply with the IRS short-low to long-low, while the factor in the column (8) is used to scale the
IRS to the MIPS 24 $\mu$m photometry. }
\end{deluxetable}

%%%%%%%%%%%%%%%%%%%%%%%%%%%%%%%%%%%%%%%%%%%%%%%%%%
\begin{deluxetable}{ccccccccccccccccc}
\tabletypesize{\scriptsize}
\tablecolumns{7}
\tablecaption{\label{data} INFRARED SPECTRA AND PHOTOMETRY OF THE SAMPLE}
\tablewidth{0pt}
\tablehead{ 
\colhead{Number}  & \colhead{Name}  &  \colhead{wave} & \colhead{$f_{\nu}$}  & \colhead{error} \\
\colhead{}        & \colhead{}      &  \colhead{}     & \colhead{[mJy]}     & \colhead{[mJy]} \\
\colhead{(1)}     & \colhead{(2)}   &  \colhead{(3)}  & \colhead{(4)}       & \colhead{(5)}  
}
\startdata
1   & PG0003+158                 & 5.52     & 7.01     & 0.43      \\
    &                            & 5.56     & 7.05     & 0.44      \\
    &                            & 5.59     & 6.95     & 0.31      \\
    &                            & 5.63     & 6.74     & 0.44      \\
    &                            & 5.67     & 6.98     & 0.42      \\
\enddata
\tablecomments{The complete version of this table is in the electronic edition of the Journal.}
\end{deluxetable}

%%%%%%%%%%%%%%%%%%%%%%%%%%%%%%%%%%%%%%%%%%%%%%%
\clearpage
\begin{landscape}
\LongTables
\begin{deluxetable}{cccccccccccccccccccccc}
%\rotate
\tabletypesize{\scriptsize}
\tablecolumns{12}
\tablecaption{\label{data_feature}  Measured Features And Physical Parameters}
\tablewidth{0pt}
\tablehead{ 
\colhead{ID}  & \colhead{Name} &  \colhead{Str$_{\rm 9.7{\mu}m-Sil}$}   &  \colhead{$\lambda^{\rm peak}_{\rm 9.7{\mu}m-Sil}$}  & 
\colhead{Str$_{\rm 18{\mu}m-Sil}$}   &  \colhead{$\lambda^{\rm peak}_{18{\mu}m-Sil}$} & \colhead{EW$_{11.3{\mu}mPAH}$}  & \colhead{Log$L_{5{\mu}m}$}    & \colhead{Log$L_{15{\mu}m}$}   & \colhead{Log$L_{18{\mu}m}$}  & \colhead{Log$L_{25{\mu}m}$}  & \colhead{log(SFR)} \\
\colhead{}  & \colhead{}  &  \colhead{}  & \colhead{[$\mu$m]} &
\colhead{}   &  \colhead{[$\mu$m]} &  \colhead{[$\mu$m]}  & \colhead{[log($L_{\odot}$)]}  &  \colhead{[log($L_{\odot}$)]}  & \colhead{[log($L_{\odot}$)]}  & \colhead{[log($L_{\odot}$)]}  & \colhead{[$M_{\odot}$/yr]} 
}
\startdata
1    & PG0003+158                    &    0.32$\pm$0.01 & 11.3  &    0.28$\pm$0.03 & 19.3  &           $<$0.032 &    11.81   &   11.74   &   11.78   &   11.63   & 1.43$\pm$0.30 \\
2    & PG0003+199                    &    0.10$\pm$0.00 & 10.3  &                  &       &    0.034$\pm$0.002 &    10.55   &   10.40   &   10.33   &   10.19   &-0.94$\pm$0.30 \\
3    & PG0007+106                    &    0.10$\pm$0.02 & 11.4  &    0.23$\pm$0.01 & 17.4  &    0.027$\pm$0.007 &    10.90   &   10.93   &   10.96   &   10.94   & 0.66$\pm$0.30 \\
4    & PG0026+129                    &    0.30$\pm$0.02 & 10.1  &    0.30$\pm$0.02 & 17.8  &           $<$0.019 &    11.24   &   10.99   &   10.99   &   10.74   &     $<$0.65   \\
5    & PG0043+039                    &    0.33$\pm$0.03 & 9.8   &    0.13$\pm$0.03 & 19.1  &           $<$0.043 &    11.75   &   11.62   &   11.66   &   11.58   & 1.64$\pm$0.30 \\
6    & PG0049+171                    &    0.30$\pm$0.03 & 10.5  &    0.53$\pm$0.03 & 17.5  &           $<$0.041 &    10.05   &   9.92    &   9.92    &   9.61    &     $<$-0.65  \\
7    & PG0050+124                    &    0.28$\pm$0.00 & 10.0  &    0.11$\pm$0.00 & 17.4  &    0.055$\pm$0.002 &    11.24   &   11.40   &   11.44   &   11.43   & 1.59$\pm$0.30 \\
8    & PG0052+251                    &    0.22$\pm$0.04 & 10.5  &    0.39$\pm$0.02 & 17.4  &    0.046$\pm$0.014 &    11.19   &   11.22   &   11.24   &   10.99   & 1.05$\pm$0.30 \\
\enddata
\end{deluxetable}
\end{landscape}

%%%%%%%%%%%%%%%%%%%%%%%%%%%%%%%%%%%%%

%%%%%%%%%%%%%%%%%%%%%%%%%%%%%%%%%%%%%%%%%%%%%%%%%%

\begin{deluxetable}{cccccccccccccccccccc}
%\rotate
\tabletypesize{\scriptsize}
\tablecolumns{15}
\tablecaption{\label{COMP_SPEC} The Median Composite Spectra Of PG and 2MASS samples}
\tablewidth{0pt}
\tablehead{ \colhead{wave[$\mu$m]}  & \colhead{log(${\nu}L_{\nu}$)} &  \colhead{log(${\nu}L_{\nu}$)}  & 
\colhead{log(${\nu}L_{\nu}$)}  & \colhead{log(${\nu}L_{\nu}$)} \\
\colhead{} & \colhead{All PG} & \colhead{Radio-quite PG} & \colhead{Radio-loud PG} & \colhead{All 2MASS} }
\startdata

5.05    & 0.035$\pm$0.151  & 0.049$\pm$0.165  & 0.013$\pm$0.096  & 0.024$\pm$0.206 \\
5.15    & 0.021$\pm$0.156  & 0.032$\pm$0.171  & -0.006$\pm$0.097 & 0.014$\pm$0.191 \\
5.25    & 0.023$\pm$0.161  & 0.024$\pm$0.174  & 0.023$\pm$0.096  & 0.008$\pm$0.246 \\
5.36    & -0.002$\pm$0.173 & -0.002$\pm$0.186 & 0.001$\pm$0.102  & 0.006$\pm$0.209 \\
5.47    & 0.001$\pm$0.159  & -0.000$\pm$0.169 & 0.004$\pm$0.101  & 0.020$\pm$0.229 \\
5.58    & -0.003$\pm$0.164 & -0.006$\pm$0.176 & -0.001$\pm$0.099 & 0.029$\pm$0.191 \\
5.69    & -0.005$\pm$0.150 & -0.019$\pm$0.160 & 0.009$\pm$0.098  & 0.025$\pm$0.150 \\
5.80    & -0.010$\pm$0.175 & -0.017$\pm$0.189 & 0.011$\pm$0.088  & -0.001$\pm$0.201\\
\enddata
\tablecomments{ }
\end{deluxetable}

%%%%%%%%%%%%%%%%%%%%%%%%%%%%%%%%%%%%%%%%%%%%%%
\begin{deluxetable}{cccccccccccccccccccc}
%\rotate
\tabletypesize{\scriptsize}
\tablecolumns{15}
\tablecaption{\label{tab_lum_lum} Correlation Between Different IR color and 15 $\mu$m  luminosity}
\tablewidth{0pt}
\tablehead{ \colhead{Y vs. X} & \colhead{A} & \colhead{B} & \colhead{$\sigma$}  }
\startdata

log($L_{25{\mu}m}/L_{15{\mu}m}$) vs. log($L_{\rm 15{\mu}m}$) & -0.19 &  0.014 & 0.14 \\
log($L_{25{\mu}m}/L_{5{\mu}m}$)  vs. log($L_{\rm 15{\mu}m}$) & -0.37 &  0.028 & 0.23 \\
log($L_{15{\mu}m}/L_{5{\mu}m}$)  vs. log($L_{\rm 15{\mu}m}$) & -0.18 &  0.014 & 0.16 \\

\enddata
\tablecomments{ Y = (A +  BX)$\pm$$\sigma$. The monochromatic luminosity is defined to be
the average emission over a 2 $\mu$m width.}
\end{deluxetable}

\end{document}